\documentclass{ws-procs961x669}            
\usepackage{ws-toc,ws-multind}
\newcommand{\apj}{ApJ}

\newcommand{\mnras}{MNRAS}

\newcommand{\aap}{A\&Ap}
\newcommand{\nat}{Nature}
\newcommand{\apjl}{ApJL}

\newcommand{\aapr}{A\&ARv}
\newcommand{\aj}{AJ}
\newcommand{\pasj}{PASJ}

\begin{document}








\wstoc{IXPE observations of supernova remnants}{R. Ferrazzoli}

\title{IXPE observations of supernova remnants}

\author{Riccardo Ferrazzoli}
\aindx{Ferrazzoli, R.}

\address{INAF Istituto di Astrofisica e Planetologia Spaziali, Via del Fosso del Cavaliere 100, 00133 Roma, Italy\\
\email{riccardo.ferrazzoli@inaf.it}}

\begin{abstract}
Supernova remnants (SNRs) are among the most important sources of non-thermal X-rays in the sky and likely contributors to Galactic cosmic rays and represent ideal targets to showcase the capabilities of the Imaging X-ray Polarimetry Explorer (IXPE) in performing spatially-resolved X-ray polarimetry.
For the first time, we can determine the turbulence level (through the measurement of polarization degree) and the orientation (through polarization direction) of the magnetic field near the shocks, where particle acceleration occurs. 
IXPE reported so far the results of the observations of four SNRs: Cas A, Tycho, SN 1006 and RX J1713.7-3946. 
These objects exhibit a wide range of characteristics, including dynamical age, spectral composition of emission, and progenitor type. 
Aptly, they revealed significantly different results among them in terms of  magnetic field properties and morphology, providing unexpected insights and shedding light on the particle acceleration mechanisms in astrophysical shocks.
\end{abstract}

\bodymatter

\begin{figure}[htbp]
	\centering
	\includegraphics[width=\textwidth]{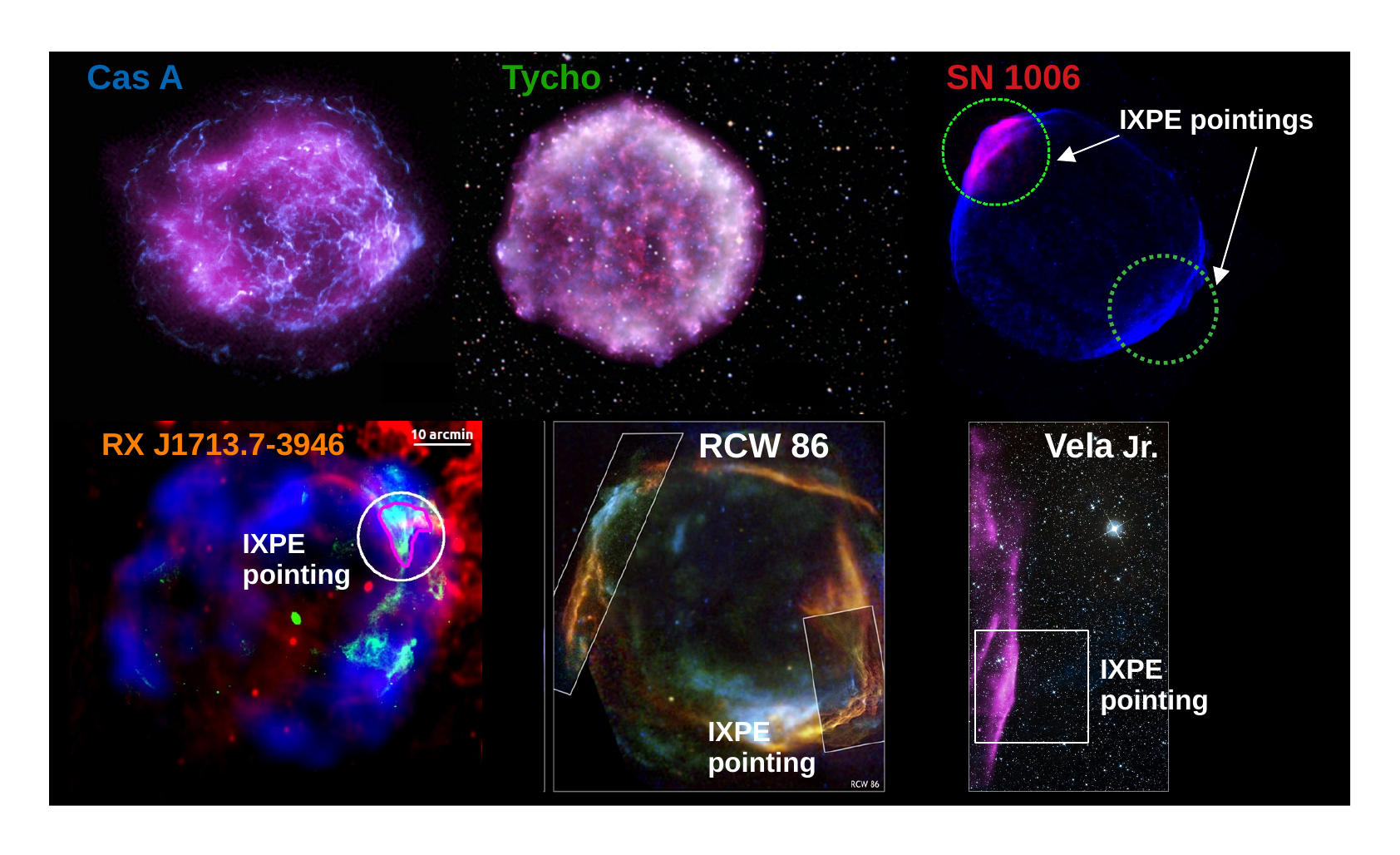}
 \caption{The six SNRs observed by IXPE. 
 The size of the objects in the images is not to-scale.
 For the most extended sources, the approximate IXPE pointing and field of view is highlighted. }
 \label{fig:Targets}
\end{figure}

\begin{figure}[htbp]
	\centering
	\includegraphics[width=\textwidth]{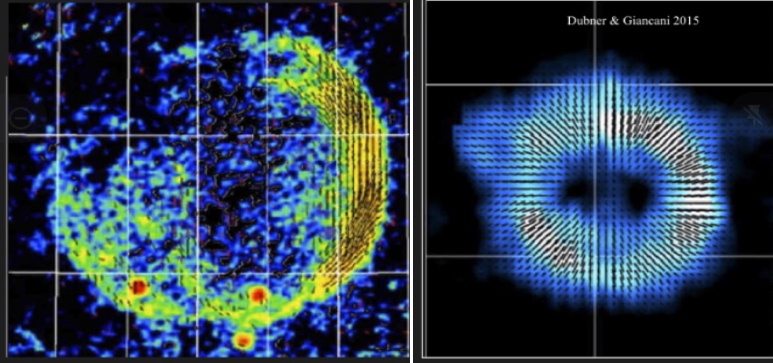}
 \caption{\textbf{Left:} magnetic field map in the radio band of the old SNR CTB01 showing a morphology that is parallel to the shock.
 \textbf{Right:} magnetic field map in the radio band of Cas A, a young SNR, whose field lines are instead radial. 
 Figures adapted from \cite{2015Dubner}.}
 \label{fig:Old_young_SNR}
\end{figure}

\begin{figure}[htbp]
	\centering
	\includegraphics[width=\textwidth]{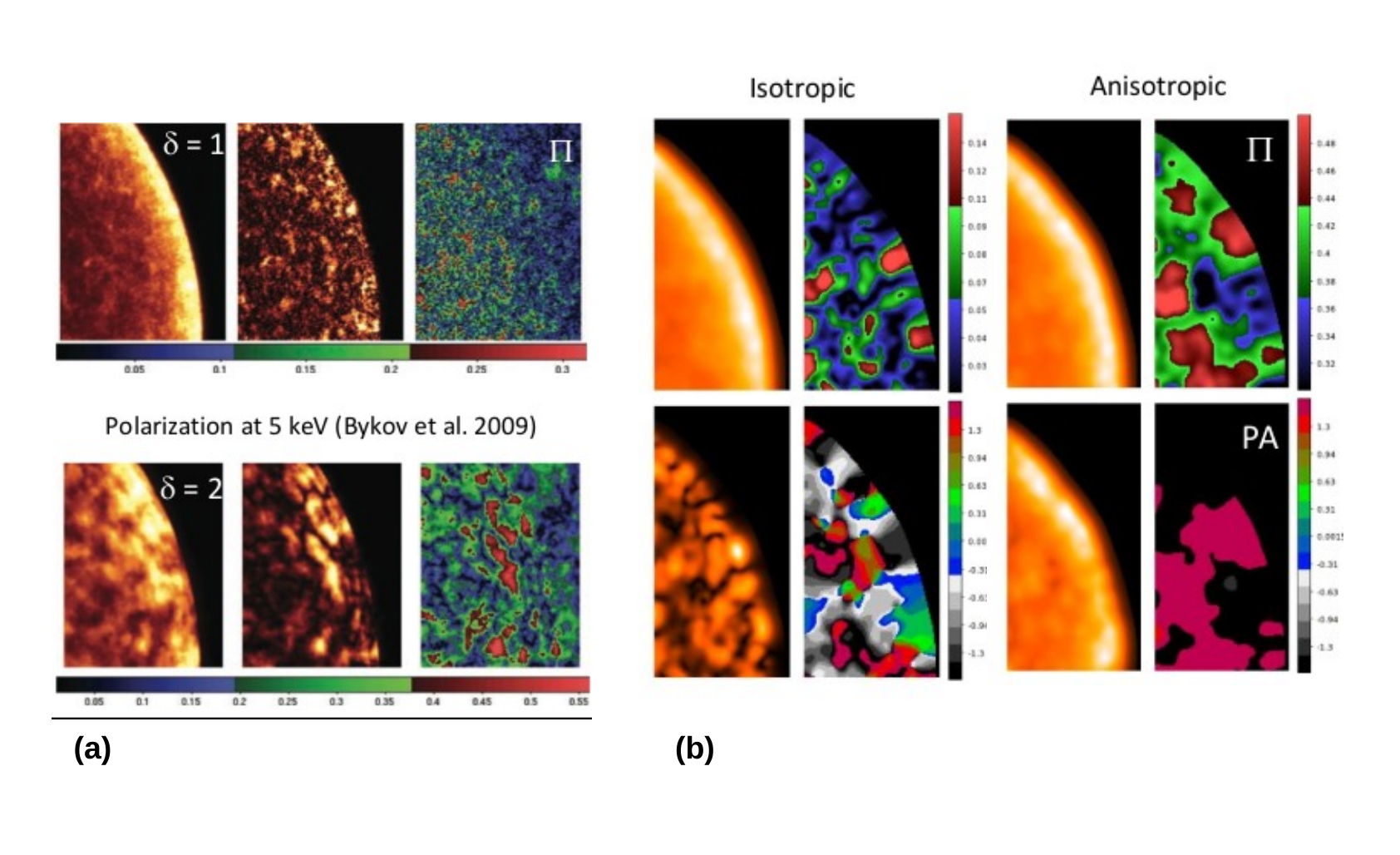}
 \caption{\textbf{Left:} simulated polarization fraction for steep (top) and flat (bottom) spectra.
 \textbf{Right:} simulated polarization fraction and angle for isotropic and anisotropic turbulence.
 Figures adapted from \cite{2009Bykov} \cite{2020Bykov}.}
 \label{fig:Bykov}
\end{figure}

Supernova remnants (SNRs) are among the extended sources that the NASA-ASI Imaging X-ray Polarimetry Explorer \cite{2022Weisskopf} is uniquely equipped to observe and study thanks to its imaging capabilities, so that for the first time we can determine not only the X-ray  polarimetric properties of these objects, but also how they change across different regions of interest. 
In its first eighteen months of operations since its launch in December 2021, IXPE observed six young SNRs: Cas A, Tycho, the north eastern and south western limbs of SN1006, the north western region of RX J1713.7-3946, the south western region of RCW86, and a rim of Vela Jr.
Of these, results from the first four have been reported so far \cite{2022Vink_b, 2023Ferrazzoli,2023Zhou,2024Ferrazzoli}.
The IXPE targets are shown in Fig.\ref{fig:Targets}.
These objects are the result of the violent death of a star, whose ejected matter interacts with the interstellar medium producing shocks.
The interest in polarization of SNR comes from the fact they are thought to be the dominant source of Galactic Cosmic Rays, through the mechanism of diffusive shock acceleration (DSA, e.g. \cite{2001Malkov}).
This efficient acceleration process requires a strong and turbulent magnetic field.
The turbulence can be either preexisting or self generated turbulence that can come from the streaming ions themselves.
Thanks to turbulence in the background plasma, particles scatter back and forth in the shocks gaining energy.
Because in many SNRs we observe in the X-rays emission from thin  filaments that we ascribe to synchrotron emission, this means that a population of relativistic electrons must exist that is accelerated very very close to the shocks.
So, even if IXPE, with its 30 arcseconds angular resolution, cannot resolve these filaments like Chandra does, we know that the X-rays, and hence the polarized emission comes from close to the shocks.
This is in contrast with, for example, radio emitting electrons that instead traveled far from the acceleration regions.
And because the synchrotron emission is intrinsically polarized, the degree of polarization we observe provides us with information about the level of order, and hence turbulence, at such small scales.
On the other hand, the polarization direction gives us information on the magnetic field, being orthogonal to the magnetic field orientation. \\
Indeed the morphology of the magnetic field of SNRs is something we are very interested in, because of an open question coming about from our knowledge of radio polarization measurements: given a shock going through an initially random magnetic field, one would expect it to compress the field component that is perpendicular to the shock motion, as a result that component is enhanced, resulting in a mostly tangential magnetic field.
In the radio band, this is exactly what it is observed in old ($ \gtrsim2000$ years) such as the SNR CTB01 shown in Fig. \ref{fig:Old_young_SNR}.
However, radio observations of  younger SNRs, such as CasA, or Tycho, radio polarization implies a radial magnetic field \cite{1976Dickel, 2015Dubner}.
This dichotomy is illustrated in Fig. \ref{fig:Old_young_SNR}.
The reason for the different morphology of the magnetic field of in young SNRs is still unknown, but many theories have been put forward and they mostly fall into two schools of thought.
The first argues for hydro-dynamical instabilities such as Rayleigh-Taylor filaments that are strung along and stretch out so that the magnetic field is carried along in those directions and the particles are spiraling along that magnetic field and that’s what is giving the radial orientation \cite{2013Inoue}.
The second calls for a sort of selection effect due to a more efficient particle acceleration in regions where the shock velocity is parallel to the magnetic field \cite{2017West}.
Hence there is a clear interest in mapping the morphology of the magnetic field in the X-rays in order to understand what the morphology of the magnetic field is closer to the shocks.
Before the launch of IXPE literature has been produced trying to predict what could have been observed depending on the characteristics of the turbulence \cite{2009Bykov,2011Bykov_a,2020Bykov}.
The expectations were that steep turbulence spectra ($\delta=2$) - indicative of have higher amplitude turbulence at low spatial scales - would lead to the possibility of observing polarization propagating downstream (Fig. \ref{fig:Bykov} (a)  bottom).
For relatively flat ($\delta=1$) spectra instead the polarization would be harder to see (Fig. \ref{fig:Bykov} (a) top).
On the other hand, the presence of built-in anisotropies in the turbulence spectrum as it is crossing the shock, depending on their scale could lead to large polarization degrees that might be resolved with IXPE (Fig. \ref{fig:Bykov} (b)).

\section{IXPE observations}

\subsection{Cas A}
\begin{figure}[htbp]
	\centering
	\includegraphics[width=\textwidth]{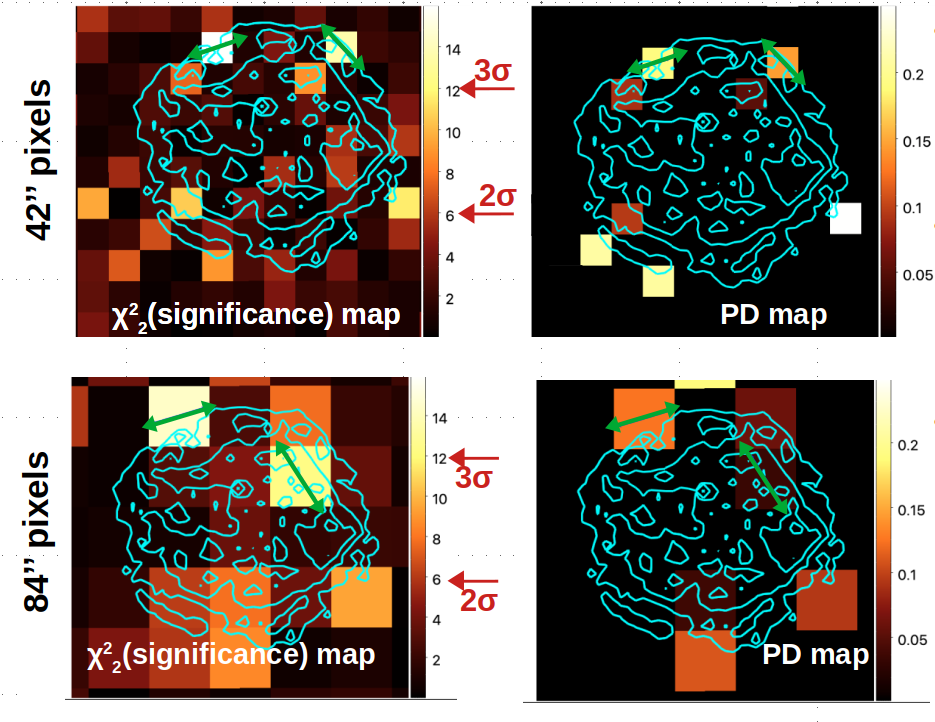}
 \caption{Figures adapted from \cite{2022Vink_b}.
 Left column: significance maps in terms of two-degrees-of-freedom $\chi^2_2$ values for the polarization signal for the 3–6 keV band. 
 Right column: the corresponding polarization degree maps. 
 Only pixels with confidence levels larger than $2 \sigma$ ($\chi^2_2 = 6.282$) are shown.
 For pixels with $\chi^2_2>11.82$ (corresponding to $3 \sigma$ confidence level) the polarization angles are indicated with green arrows. 
 The errors on these angles are $\sim 8$ degrees. 
 Top row: maps with pixel sizes binned to 42". 
 Peaks in the $\chi^2_2$ map are 15.9, 13.62 corresponding to polarization degrees of $19\%$ and $14.5\%$, respectively. 
 Bottom row: same plot, but with a larger pixel size of 84". 
 Peaks in the $\chi^2_2$ map are 14.4, 12.32 corresponding to polarization degrees of $12.4\%$ and $3.4\%$, respectively. }
 \label{fig:CasA_maps}
\end{figure}

\begin{figure}[htbp]
	\centering
	\includegraphics[width=\textwidth]{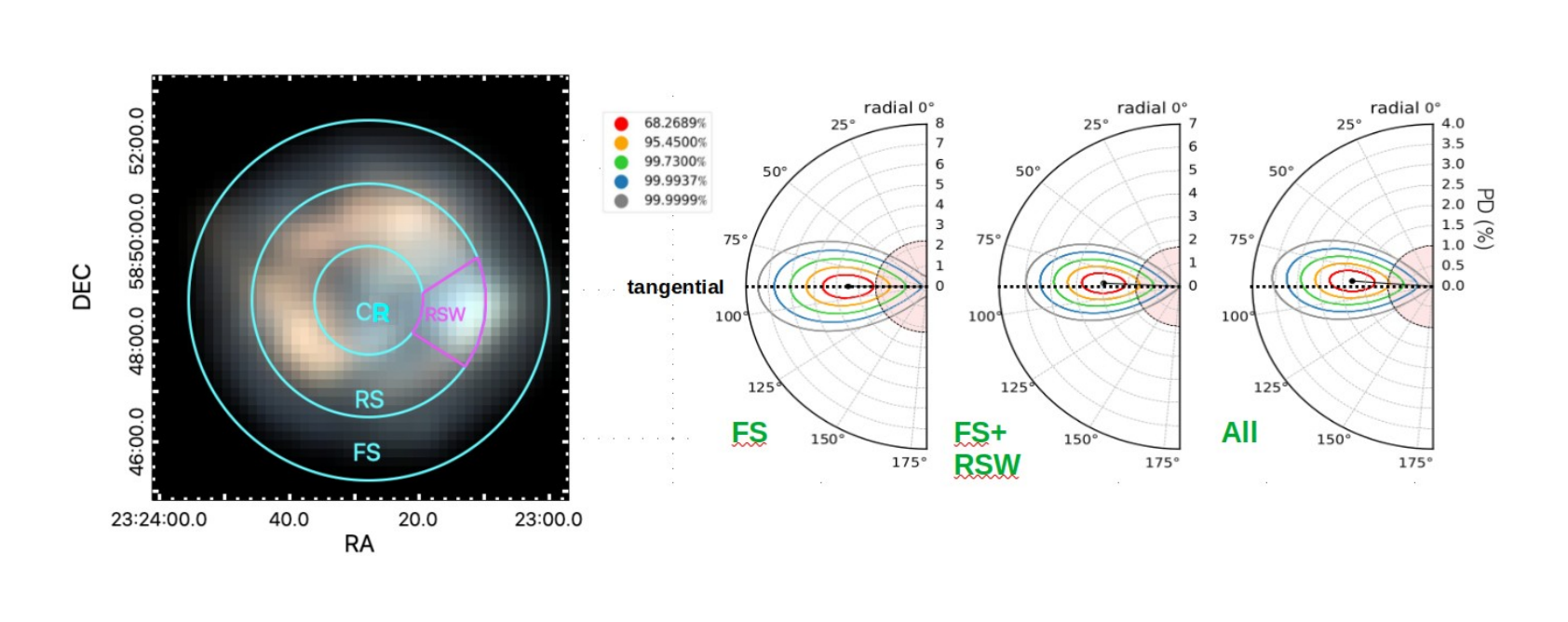}
 \caption{
 \textbf{Left:} IXPE three color Stokes I image, based on the 2–3 keV, 3–4 keV, and 4–6 keV bands, combined from the three detectors, with superimposed regions that were used to test for an overall radial or tangential polarization vector orientation. 
 The regions of interest used for the Stokes parameter alignment analysis are tagged as Central Region (CR), Reverse Shock (RS), Reverse Shock West (RSW, in magenta), and Forward Shock (FS). Figure adapted from \cite{2022Vink_b}.
 \textbf{Right:} Polar plots of the the measured polarization degree and angle with respect to circular symmetry as confidence contours for the Forward Shock (FS), combination of Forward Shock and Reverse Shock West (FS+RSW), and for the whole remnant (All). 
 The radial coordinate indicates the polarization degree in percent. 
 The pink shaded region corresponds to the MDP99 level. 
 Values compatible with 90° correspond to an overall tangentially oriented polarization averaged over the region, while around 0° indicates on average a radially oriented polarization.}
 \label{fig:CasA_regions_and_polarplots}
\end{figure}

The SNR Cas A was the first target of the scientific campaign after the launch and commissioning, and it represented the benchmark for all the analysis techniques developed by the IXPE collaboration.
Cas A is a bright and young ($\sim350$ years old) relic of a core-collapse supernova whose distance is 3.4 kpc \cite{1995Reed}.
Expansion measurements establish 
Its forward shock velocity of $\sim 5000{\rm\ km\ s}^{-1}$ \cite{1998Vink,2009Patnaude}. 
In the radio band, Cas A is observed to have a radially oriented magnetic field\cite{1970Rosenberg, 1987Braun} and an average 5\% polarization degree that arrives up to 8-10\% in the outer rim \cite{1995Anderson}. 
IXPE observed Cas A for $\sim 900$ ks, and the results were presented in \cite{2022Vink_b}. \\
A pixel-by pixel search of polarization signal, considering spatial bins of size 42'' and 84'' (see Fig. \ref{fig:CasA_maps}) revealed weak evidence for X-ray polarization of $\rm PD =5 - 15$\%, and tangential PA, suggesting radial magnetic fields. 
In order to improve the sensitivity, circular symmetry of the polarization direction can be assumed, and exploiting the additivity of the Stokes parameters larger emission regions could be combined.
This method established highly significant detections of polarization from multiple regions: the forward shock, the forward shock and a distinct region on the western edge of the reverse shock, and the whole remnant.
This is illustrated in Fig. \ref{fig:CasA_regions_and_polarplots} that shows a three-color IXPE image of Cas A with these regions identified (left), and their polarization plots (right).

In the polarization plots the PD is represented by the radial component of the polar plot and the PA is measured relative to the radial direction. 
Contours show confidence intervals, and the pink circle illustrates the minimum detectable polarization at 99\% confidence. 
The PA values for all regions are consistent with tangential polarization -- and thus radial magnetic fields -- with PD values ranging from $\sim 2-4$\%. 
Correcting for the dilution due to the thermal flux, determined using Chandra spectral modeling of the regions of interest, the PD values of the X-ray synchrotron emission alone are in the range of $\sim 2 - 5$\%, lower than observed in the radio band.

The low PD measured is suggestive of high levels of turbulence in the region close to the shock.
The measured PA instead implies that that whatever the process that is responsible for the radial magnetic field observed in the radio, is already at work very close to the shock where particle are accelerated, on spatial scales corresponding
to the thin rims ($\sim 10^{17}$~cm.

\subsection{Tycho}
\begin{figure}[htbp]
	\centering
	\includegraphics[width=\textwidth]{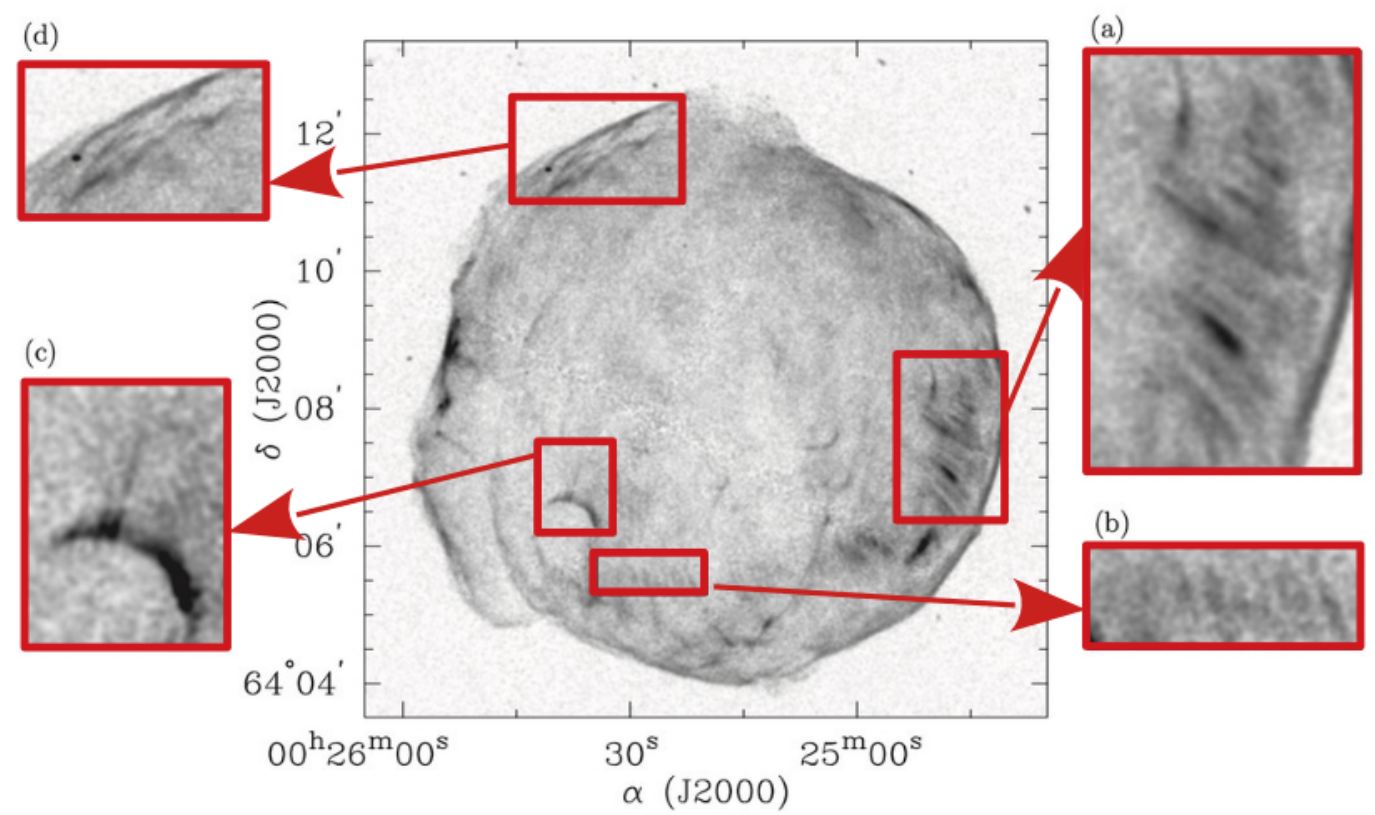}
 \caption{Figure adapted from \cite{2011Eriksen}.
 Chandra X-ray 4.0–6.0 keV image of the Tycho SNR, showing various regions of striping in the nonthermal emission. 
 Clockwise from the upper right: (a) the western stripes; (b) a fainter southern
stripes; (c) arch of nonthermal emission;  (d) north-eastern filaments.
}
\label{fig:Tycho_eriksen}
\end{figure}

\begin{figure}[htbp]
	\centering
	\includegraphics[width=\textwidth]{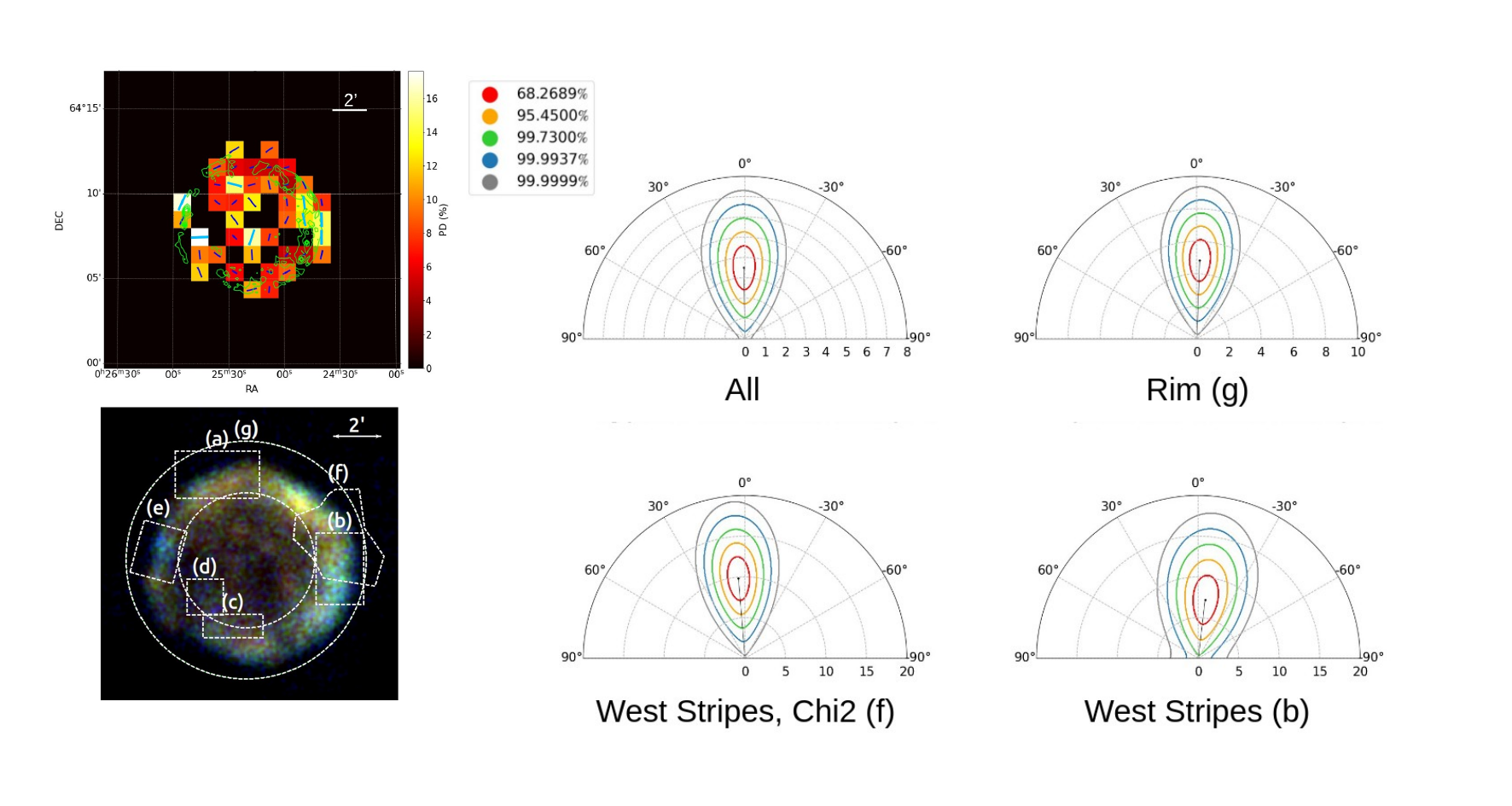}
 \caption{Figures adapted from \cite{2023Ferrazzoli}.
 \textbf{Top left:} polarization map in the 3–6 keV energy band with a 60" pixel size. Only the pixels with significance higher than 1$\sigma$ are shown. The blue bars represent the polarization direction (that is, the direction of the electric vector polarization angle) and their length is proportional to the polarization degree. The thicker cyan bars mark the pixels with significance higher than 2$\sigma$. The orientation of the magnetic-field is perpendicular to the polarization direction. Superimposed in green are the 4–6 keV Chandra contours.
 \textbf{Bottom Left:} regions of interest.
 \textbf{Right:} polar plots for the most significant Tycho regions of interest. 
 Each diagram depicts the measured polarization degree, and direction with respect to circular symmetry with respect to the geometrical center of the remnant, as confidence contours. 
 The confidence levels are given color-coded in the legend. 
 The radial coordinate indicates the polarization degree in percent. 
 Values more consistent with a polarization direction of 0° correspond to an overall tangentially oriented polarization averaged over the region.
}
\label{fig:Tycho}
\end{figure}

Unlike Cas A, a core-collapse remnant, Tycho is the result of a Type Ia explosion observed as the historical supernova SN 1572 \cite{2003Green}.
Its shock velocity lies between 3500 and 4400 km s$^{-1}$ \cite{2017Williams, 2020Williams}.
Polarization has been detected in the radio band at 2.8--6 cm, with PD values ranging from 0 at the center, to 7--8\% at the outer rim, with the PA indicating a large-scale radial magnetic-ﬁeld structure \cite{1971Kundu,1973Strom, 1975Duin, 1991Dickel, 1997Reynoso}. 
A bright thin synchrotron rim at the shock was revealed by high resolution Chandra X-ray observations \cite{2002Hwang}, as well as unique small-scale structures (from $\sim$arcseconds to $\sim$arcminute) in the 4–6 keV range, known as "stripes" in the western rim, first noted by \cite{2011Eriksen} (see Fig. \ref{fig:Tycho_eriksen}). 
The stripes, along with the SNR rim, are considered possible sites for cosmic-ray acceleration and may result from rapid energy losses of TeV electrons emitting X-rays downstream of the shock in enhanced magnetic fields \cite{2020Bykov}. 
If this holds true, the synchrotron structures arise from the geometric projection of the thin regions where the TeV electrons are accelerated, and their polarization properties have been widely discussed leading up to the IXPE mission \cite{2009Bykov,2011Bykov_a,2020Bykov}.
%
Tycho was the second SNR observed by IXPE, during the Summer of 2022, for a total exposure time of $\sim990$ ks \cite{2023Ferrazzoli}. 
The analysis followed a similar approach to that used for Cas A, beginning with a pixel-by-pixel search for a signal.
However, since Tycho is not as bright as Cas A, the polarization map binned on a 1 arcminute scale does not show highly significant detections.
By aligning and summing over the data from different regions of interest, such as the highest significant region in the west, the rim, but also the whole remnant as shown in the left panel of Fig. \ref{fig:Tycho}, highly significant detections of polarized emission were identified.
The measured tangential direction of polarization corresponds to a radial magnetic ﬁeld, consistent with the radio band polarization observations but originating from regions even closer to the shock.
The degree of X-ray polarization in Tycho is significantly higher than for Cas A: in Tycho the synchtrotron PD was found to be $12\%\pm2\%$ in the rim and $9\%\pm2\%$ in the whole remnant. 
These results are compatible with the expectation of turbulence produced by an anisotropic cascade of a radial magnetic ﬁeld near the shock.

\subsection{SN 1006}
\begin{figure}[htbp]
\includegraphics[width=\textwidth]{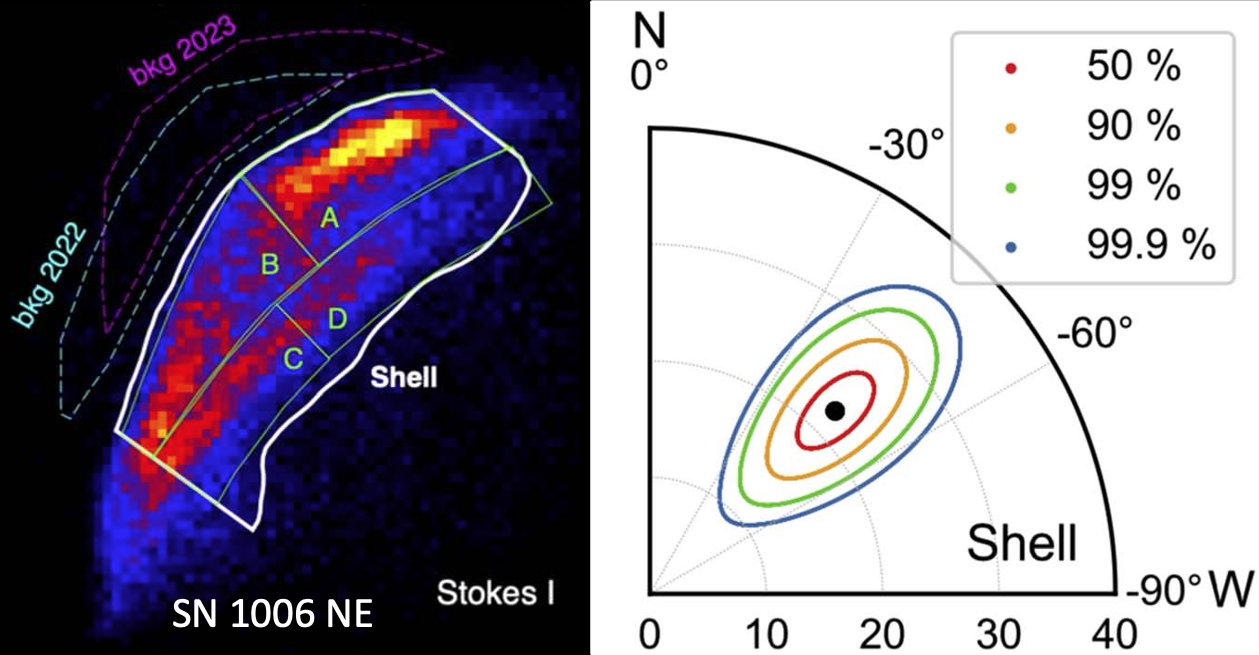}
\caption{\textbf{Left:} IXPE image of the north eastern limb of SN1006. 
The sub-regions of interest are the highlited and labelled in green while the white contour  indicates the entire visible part of the shell. 
Labels identify regions investigated in \cite{2023Zhou}. 
\textbf{Right:} Polar plot for entire shell region.
Figures adapted from \cite{2023Zhou}}.
\label{fig:SN1006}
\end{figure}

SN 1006 was the first supernova remnant (SNR) detected to emit strong nonthermal X-rays \cite{1995Koyama}, making it an ideal target for X-ray polarimetry. 
IXPE observed the northeastern shell of SN 1006 (SN 1006 NE) twice from August 5–19, 2022 and March 3–10, 2023, with a total effective exposure of around 960 ks \cite{2023Zhou}. 
Compared to Cas A and Tycho, SN 1006 is significantly more extended, but fainter in X-rays, hence background contributions from the source region were substantial, requiring their removal during analysis.
IXPE resolved the double-rim structure of SN 1006 NE, enabling a detailed spatial analysis of the X-ray polarization. 
The Stokes I image of SN 1006 NE in the 2–4 keV range is shown in Fig. \ref{fig:SN1006}, with the white and green polygon regions indicating the shell and four sub-scale areas, respectively. 
Polarization of the entire shell was detected at a significance of $6.3\sigma$, with a polarization degree of $22.4\pm 3.5\%$ and a polarization angle of $-45.5^\circ$, indicative of a radial magnetic field also in this case. 
The sub-regions showed no significant variations of the polarization properties among them.

\subsection{RX J1713.7-3946}
\begin{figure}[htbp]
\includegraphics[width=\textwidth]{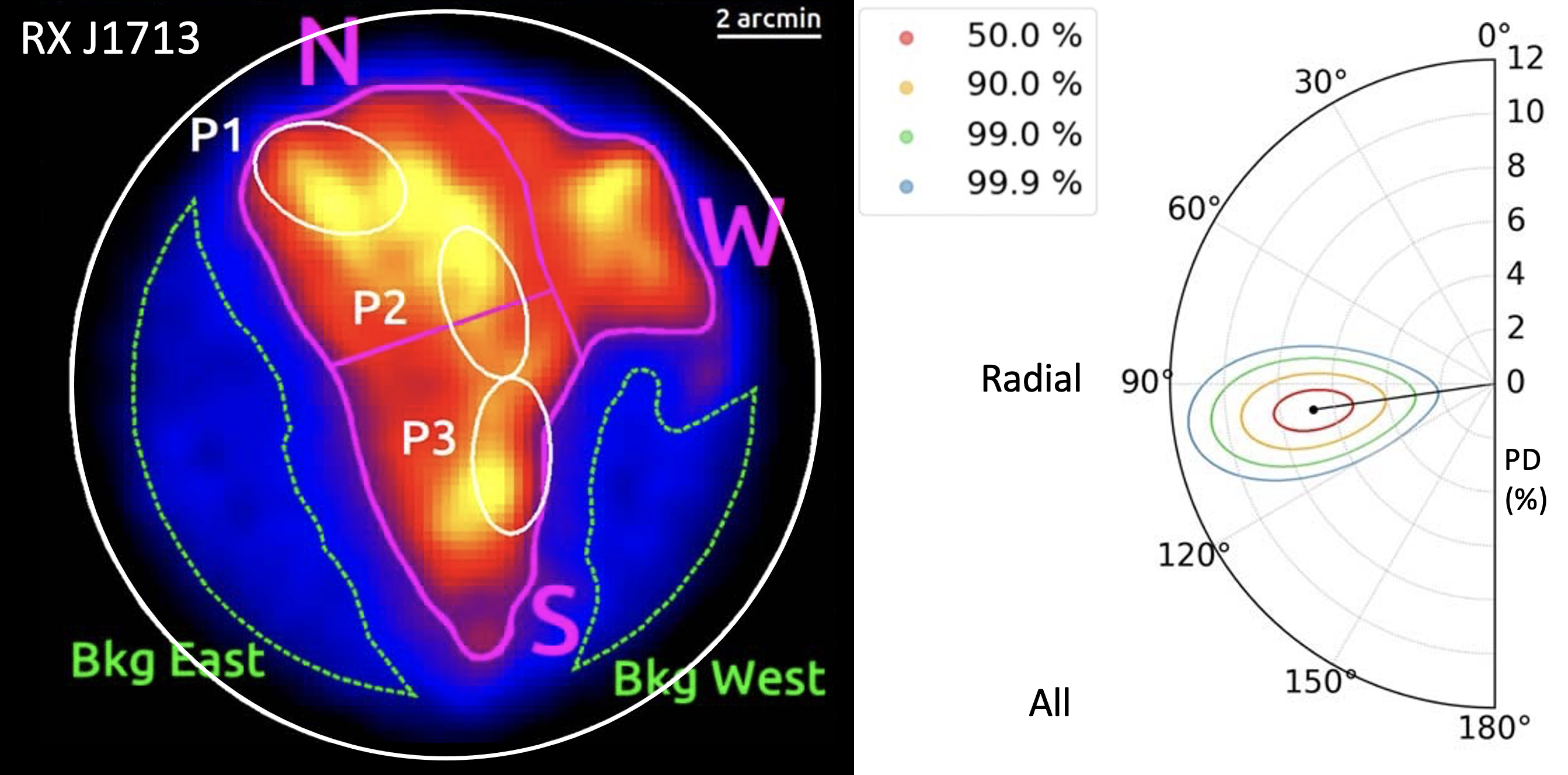}
\caption{\textbf{Left:} IXPE image of the north western region of RX J1713.7-3946. 
Labels identify the regions considered in \cite{2024Ferrazzoli}. 
\textbf{Right:} Polar plot of the results for the entire region.
Figures adapted from \cite{2024Ferrazzoli}}
\label{fig:RXJ1713}
\end{figure}

RX J1713.7-3946 is a large shell-type supernova remnant (SNR), approximately 1 degree in diameter, located in the Galactic plane. 
It was first identified in the ROSAT all-sky survey \cite{1996Pfeffermann}. 
Situated around 1 kpc away \cite{2003Fukui,2003Uchiyama,2004Cassam-Chenai}, it is believed to have resulted from a Type Ib/c supernova, potentially linked to the historical SN 393 event \cite{1997Wang,2016Tsuji,2017Acero}, making it the oldest SNR whose results have been reported by IXPE to date. 
Similarly to SN1006, its hard X-ray emission is exclusively nonthermal \cite{1997Koyama,1999Slane}.
The shock velocity in the northwestern (NW) region was measured at approximately 3900 km/s, with slower velocities down to 1400 km/s in other structures within the NW shell \cite{2016Tsuji}. 
Earlier radio polarization measurements detected signals in the NW shell, though Faraday rotation prevented determination of the magnetic field direction \cite{2004Lazendic}. 
RX J1713.7-3946 is also well-studied in the $\gamma$-ray band, with ongoing debates about whether its high-energy emissions are of hadronic or leptonic origin.
IXPE conducted three observations of the NW shell of RX J1713.7-3946 in 2023: from August 24-27, August 28 to September 1, and September 24 to October 5, totaling around 841 ks of exposure \cite{2024Ferrazzoli}.
The left panel of Fig. \ref{fig:RXJ1713} shows the regions of interest selected for analysis and the background extraction areas. 
Analysis of both the polarization map and regions of interest revealed that the polarization direction is perpendicular to the shock, with an average polarization degree of $12.5\pm3.3\%$ across the entire region. 
Unlike other IXPE-observed remnants, which exhibit radial magnetic fields, RX J1713.7-3946 is the first case where shock-compressed tangential magnetic fields dominate in the X-ray band. These results align with a model where shock compression of upstream isotropic turbulence generates a primarily tangential magnetic field.

\section{Discussion and conclusions}
The current picture of the X-ray polarimetry of SNR shows significant detection of polarization in all four sources: Cas A, Tycho, SN1006, and RX J1713.7-3946.
The synchrotron PD is in the range of 5\% – 30\%, implying high level of turbulence, as expected for efficient particle acceleration.
However, values vary significantly between remnants, suggesting inherent properties that drive instability amplitudes and scales.
The observation of radial magnetic field very near shock of Cas A, Tycho, and SN1006 presents constraints on the formation of instabilities and the cascading of anisotropic turbulence in fast shocks.
One exception - so far - is the detection of a tangential magnetic field in RX J1713.7-3946, that is consistent with shock compression of isotropic (or partially-tangential) upstream field, with little development of radial field in downstream region.
In Table \ref{tab:comparison} the IXPE polarimetric results and the physical characteristics of each source are listed.
Because of the limited sample it is difficult to infer a trend.
\begin{table}[htbp]
\caption{Physical and polarization parameters parameters for the SNRs observed by IXPE.}
\footnotesize
\begin{center}
\begin{tabular}{lccccccc}
\toprule
\toprule
\noalign{\smallskip}
\noalign{\smallskip}

            & $V_s$         & $n_0$           & $\eta$      & $\rm B_{low}$ & PD$_{Rim}$       & PD$_{SNR}$      & PD$_{Peak}$\\
            & (km s$^{-1}$) & cm$^{-3}$       &             & ($\mu$G)     & (\%)           & (\%)          & (\%)        \\
\hline
Cas A       & $\sim 5800$   & $0.9 \pm 0.3$   & $\sim 1-6$  & $25-40$      & $4.5 \pm 1.0$  & $2.5 \pm 0.5$ & $\sim 15$   \\
Tycho       & $\sim 4600$   & $\sim 0.1-0.2$  & $\sim 1-5$  & $30-40$      & $12 \pm 2$     & $9 \pm 2$     & $23 \pm 4$  \\
SN~1006 NE  & $\sim 5000$   & $\sim0.05-0.08$ & $\sim 6-10$ & $18-26$      & $22.4 \pm 3.5$ & ...           & $31 \pm 8$  \\
RX J1713 NW & $1400-2900$   & $\sim0.01-0.2$  & $\sim 1.4$  & $\sim 10$    & $13.0 \pm 3.5$ & ...           & $46 \pm 10$ \\
\hline
\end{tabular}
\label{tab:comparison}
\end{center}
\end{table}
There is a possible connection with the Bohm factor $\eta$, that is a measure of the acceleration efficiency, or with the ambient density $n_0$, which can be related with the ability to generate the turbulence that propagates downstream.
The observation of SNRs has been the perfect showcase of the IXPE capabilities and allowed us to learn about the turbulence and magnetic field morphology of these objects at unprecedented scales.
Future results coming from the observation of Vela Jr., RCW86, and the south western limb of SN1006 will help to improve our knowledge of these sources.

\section*{Acknowledgments}
The Imaging X-ray Polarimetry Explorer (IXPE) is a joint US and Italian mission.  
The US contribution is supported by the National Aeronautics and Space Administration (NASA) and led and managed by its Marshall Space Flight Center (MSFC), with industry partner Ball Aerospace (contract NNM15AA18C).  
The Italian contribution is supported by the Italian Space Agency (Agenzia Spaziale Italiana, ASI) through contract ASI-OHBI-2022-13-I.0, agreements ASI-INAF-2022-19-HH.0 and ASI-INFN-2017.13-H0, and its Space Science Data Center (SSDC) with agreements ASI-INAF-2022-14-HH.0 and ASI-INFN 2021-43-HH.0, and by the Istituto Nazionale di Astrofisica (INAF) and the Istituto Nazionale di Fisica Nucleare (INFN) in Italy.  
This research used data products provided by the IXPE Team (MSFC, SSDC, INAF, and INFN) and distributed with additional software tools by the High-Energy Astrophysics Science Archive Research Center (HEASARC), at NASA Goddard Space Flight Center (GSFC).
R.F. is partially supported by MAECI with grant CN24GR08 “GRBAXP: Guangxi-Rome Bilateral Agreement for X-ray Polarimetry in Astrophysics”.

\vfill
\pagebreak



\end{document}